\documentstyle[preprint,aps]{revtex}
\newcommand{\be}{\begin{eqnarray}}
\newcommand{\ee}{\end{eqnarray}}

\newcommand{\zbb}{Z \rightarrow b \bar{b}}
\newcommand{\non}{nonuniversal }
\newcommand{\as}{\alpha_s}
\newcommand{\xx}{\chi^2}

\begin{document}
\preprint{
KAIST-CHEP-95/14
}
\title{
$\alpha_s(M_Z^2)$ and $R_b$ discrepancy
with nonuniversal interactions
}
\vspace{0.65in}
\author{
Jae Kwan ~Kim,
Yeong Gyun ~Kim\thanks{ygkim@chep6.kaist.ac.kr},~
Jae Sik ~Lee\thanks{jslee@chep6.kaist.ac.kr}~
and~
Kang Young ~Lee\thanks{kylee@chep5.kaist.ac.kr}
}
\vspace{.5in}
\address{
Department of Physics, KAIST, Taejon 305--701, KOREA}
\date{\today}
\maketitle
\begin{abstract}
We implement global fits to LEP data with the \non interactions.
Consistent $R_b$ with experimental value and
consistent $\as(M_Z^2)$ with that from low energy experiments
are obtained.
We also find that the $\xx$ is better than the Standard Model.
And we argue that other kinds of new physics are needed
to explain the difference between the values of $\as(M_Z^2)$
from low energy experiments and from the 3-jet ratio.
\end{abstract}
\pacs{ }

\narrowtext


Recently the Collider Detector of Fermilab (CDF) Collaboration
presented evidence for a top quark with a mass $m_t \sim$175 GeV
\cite{cdf}.
Such a heavy top affects the partial width of $\zbb$
and recent analysis indicates that
the experimentally measured value for the ratio
$R_b \equiv \Gamma(\zbb)/\Gamma(Z \to \mbox{hadron})$
is higher than the Standard Model (SM) prediction
at a 2.5$\sigma$ level \cite{lep,ewgroup}.
This discrepancy may be the first signal for new physics
beyond the SM if it will be confirmed by future experiments.
A number of possible scinarios of new physics are being
suggested to explain this $R_b$ discrepancy.

The \non interaction acting on only the third generation
attracts us as a candidate for the new physics
since the SM predictions for other flavours should not
be disrupted by the new physics.
Models of this type are motivated by the idea
that the top quark has a mass of the order of the weak scale
and would play a key--role in electroweak symmetry breaking
via top quark condensation \cite{non1}.
Considering the general approach, the anomalous \non
interaction terms are SU(2)$_L$ $\times$ U(1)$_Y$ invariant and
the $b$--quark will take part in top quark interactions
when the left--handed doublet is involved.
This can result in a modification of the $\zbb$ vertex.
We parametrize the \non interaction effects in the $\zbb$ vertex
by introducing the parameters $\kappa_{L,R}$.
These parameters shift the SM tree level couplings of
the neutral currents $g_{L,R}$ to the effective couplings
$g_{L,R}^{\mbox{\footnotesize{eff}}}$
\be
g_{L,R}^{\mbox{\footnotesize{eff}}}= g_{L,R} (1+\kappa_{L,R})
\ee
where
\be
g_L = -\frac{1}{2} + \frac{1}{3} \sin^2 \theta_W,~~~~
g_R = \frac{1}{3} \sin^2 \theta_W~~~.
\nonumber
\ee
It was shown that $R_b$ could be fitted to the LEP data
within 1$\sigma$ with \non interactions
in the ref. \cite{non2}.

Another possible deviation of the LEP/SLC data from the SM
is being proposed.
Shifman \cite{shifman} has pointed out that
the value of the strong coupling constant
$\as(M_Z^2) \simeq 0.126$ determined
by global fits to the $Z$--line shape variables
at the $Z$--peak
shows much discrepancy with $\as(M_Z^2) \simeq 0.112$
extracted from
low energy experiments, which is scaled to $M_Z$ scale.
And we note that the value of $\as(M_Z^2) \simeq 0.119$
from events shape variables also shows difference from
that from low energy experiments.
Kane et al. \cite{kane} noticed this point
in relation to the $R_b$ discrepancy.
They reanalyzed the LEP/SLC data
in minimal supersymmetric standard model (MSSM) scheme
with light superpartners
and found that the global fit with low $\as(M_Z^2) = 0.112$
yields better fit to the data than that of the SM.
Several authors have noted that if $R_b$ is explained by new physics,
then in general $\as(M_Z^2)$ will decrease.

In this paper we study the model with \non interactions
to explain the $\as$ problem and $R_b$ discrepancy.
We do not construct a specific model but use the effective
lagrangian technique.
We take the $\zbb$ vertex to be given phenomenologically by
the expression
\be
{\cal L} \sim Z^{\mu} ( \bar{b} \gamma_{\mu}
       (g_{V}^{\mbox{\footnotesize{eff}}}+g_{A}^{\mbox{\footnotesize{eff}}}
\gamma_5) b)
\ee
where $g_{V}^{\mbox{\footnotesize{eff}}}$ and
$g_{A}^{\mbox{\footnotesize{eff}}}$ are
 the effective vector and axial coupling constants given by
\be
g_{V}^{\mbox{\footnotesize{eff}}}&=&2(g_{R}^{\mbox{\footnotesize{eff}}}+g_{L}^{\mbox{\footnotesize{eff}}})
\nonumber \\
g_{A}^{\mbox{\footnotesize{eff}}}&=&
2(g_{R}^{\mbox{\footnotesize{eff}}}-g_{L}^{\mbox{\footnotesize{eff}}})
{}~~~.
\ee

We used ZFITTER \cite{zfit} with the function minimizing
program MINUIT \cite{minuit} to perform the $\xx$ fit
for the LEP observables.
By $\xx$ fitting to the LEP observables with \non interactions,
we find that the value of $\as(M_Z^2) = 0.103$
lies at the global $\xx$ minimum.

Alternatively we consider the extraction of $\as$ from 3--jet ratio.
We observe that this jet variable is very insensitive to
the modification of the $\zbb$ vertex given in eq. (1).
So we find that this jet variable can be used to
extract $\as(M_Z^2)$ independetly of such kinds of new physics
that effectively change $g_L$ and $g_R$.


For completeness, we implement the $\xx$ fit to the data
in the SM framework at first.
The SM value of $\as(M_Z^2)$ from the $Z$ line shape variables
has been reported to be $\as(M_Z^2)$=0.126$\pm$0.005
by LEP Electroweak working group \cite{ewgroup}.
We use the set of following 12 variables
in our fitting procedure \cite{lep}:
$M_W,~\Gamma_Z,~\sigma_{tot},
{}~R_l \equiv \Gamma_{had}/\Gamma_{lepton},
{}~A_{FB}^{lep},~A_{\tau},~A_e,~R_b,~R_c,~A_{FB}^b,~A_{FB}^c,
{}~\sin^2 \theta_W^{lep}$.
The Higgs mass is fixed to be 100 GeV.
Our $\xx$ fit is not sensitive to the values of Higgs mass
in the region $m_H$=100--1000 GeV.
As fitting parameters, we use $t$--quark mass $m_t$ and $\as$.
We obtain followings:
\be
m_t=162.67\pm8.98 \mbox{GeV}, ~~~~~\as(M_Z^2)=0.121\pm0.004~~~.
\nonumber
\ee
These results are consistent with the fits obtained by the LEP
Electroweak working group.


The deviation of $\Gamma_b$ from the SM by the effects of
$\kappa_{L,R}$ is expressed by
\be
\frac{\delta \Gamma_b}{\Gamma_b} \sim
         2 \frac{g_L^2 \kappa_L+g_R^2 \kappa_R}
                {g_L^2 +g_R^2}~~~.
\ee
Since $g_L^2 \gg g_R^2$,
$\kappa_R$ does not affect much on $\Gamma_b$ and
we can neglect the second term.
Therefore we fix $\kappa_R=0$ in our analysis.

With a nonzero parameter $\kappa_L$,
we implement the $\xx$ fit to the same set of LEP observables.
We found the much better $\xx$ than the SM,
well--agreed $R_b$ within 1$\sigma$ range
of experimentally measured value
and the lower $\as(M_Z^2)$ than that of the SM.
We obtain the values:
\be
m_t&=&165.33 \pm 8.70~~ \mbox{GeV}~~~,
\nonumber \\
\as(M_Z^2)&=&0.103 \pm 0.009~~~,
\nonumber \\
\kappa_L&=&0.013 \pm 0.006~~~.
\nonumber
\ee
The results of our $\xx$ fit to LEP observables
are summarized in Table 1 compared with those of the SM.
In Fig. 1, we plot $R_b$ as a function of $\kappa_L$
for these values of $m_t$ and $\as(M_Z^2)$.

Because we take a model--independent approach,
we do not explicitly describe the parameter $\kappa_L$
by specific physical quantities here.
We know, however, that $\kappa_L$ is related to
the new physics scale $\Lambda$.
For example, if we take the relevant term of the effective
lagrangian as the 4--fermion coupling
\be
{\cal L}_{\mbox{\footnotesize{eff}}} \sim -\frac{1}{\Lambda^2}
               \bar{b} \gamma_{\mu} b
             ~ \bar{t} \gamma^{\mu}
               (g_V+g_A \gamma_5) t~~~,
\ee
$\kappa_L$ is computed by $t$--quark correction to the $\zbb$ vertex
as follows
\be
\kappa_L = \frac{g_A}{g_L} \frac{N_c}{8 \pi^2} \frac{m_t^2}{\Lambda^2}
           \ln \left( \frac{\Lambda^2}{m_t^2} \right)~~~.
\ee
Our fit result $\kappa_L \sim 0.013$ yields
$\Lambda \sim 1.5$ TeV from eq. (6).


The value of $\as(M_Z^2)$ at the Z--peak can also be
extracted from jet event shape variables.
There are several jet variables; thrust, jet mass,
energy--energy corelation, oblateness, C--parameter, jet multiplicity
and 3--jet ratio etc..
Here we explore the effects of the \non interactions on
3--jet ratio and determnation of $\as(M_Z^2)$.

Jets are defined as a bunch of particles
based on jet--clustering algorithms.
For example, with a jet--clustering algorithm
in the EM scheme \cite{jet},
two particles are regarded as belonging to the same jet
if their momenta satisfy the condition
\be
y_c > y_{ij} = 2 \frac{p_i \cdot p_j}{s}
\ee
where $\sqrt{s}$ is the total energy of collision
and $y_c$, so--called $y$--cut, is a given resolution parameter.

We used the 3--jet decay width at the $Z$ peak formula
derived by Bilenky et al.,
which is calculated up to the order of $\as$
and $r_b \equiv m_b^2/m_Z^2$.
Their analytic expressions are found in ref. \cite{bilenky}.
We calculate the ratio of $\Gamma_{3jet}^b$ to $\Gamma_b$
with the \non interactions given in eq. (1)
for the values of the parameter $\kappa_L$=0, 0.02, 0.08.
\be
R_{3j} &=& 0.2450~~~~~~~~\mbox{for}~~~\kappa_L = 0~~~,
\nonumber \\
R_{3j} &=& 0.2449~~~~~~~~\mbox{for}~~~\kappa_L = 0.02~~~,
\nonumber \\
R_{3j} &=& 0.2448~~~~~~~~\mbox{for}~~~\kappa_L = 0.08~~~.
\ee
We used $\as = 0.119$ which is reported
by LEP Electroweak working group for event shape variables
\cite{ewgroup}.
Each value of $\kappa_L$ corresponds to
the Standard Model, $\Lambda \sim 1$ TeV and
$\Lambda \sim 300$ GeV if we assume the effective lagrangian
such as eq. (5).
The change of 3--jet ratio with varying $\kappa_L$
is very slight and
it cannot change the value of $\as(M_Z^2)$.
We conclude that this variable is very insensitive to
the change of the parameter $\kappa_L$
and the value of $\as(M_Z^2)$ extracted from this variable
is not lowered by introduction of the new physics effects such as eq. (1),
contrary to the case of the line shape variables.


When one introduce the new physics beyond the SM
to cure the $R_b$ discrepancy,
the value of $\as(M_Z^2)$ is usually known to be
lower than that extracted from the SM.
This fact can be the answer of the problem
that the value of $\as(M_Z^2)$ emerging from the global fits
on the data at the $Z$--peak is almost 3$\sigma$ deviations
higher than the value stemming from the low energy phenomenology.
With a generic \non correction given in eq. (1),
we implemented the global fits to the observables of LEP
and found that $\as(M_Z^2) \simeq 0.103$ gives the best fit.
All the data including $R_b$ are consistent with our model predictions.

We also found that the 3--jet ratio is very insensitive
to this \non correction.
If we predict this jet variable more exactly,
therefore, we can extract $\as(M_Z^2)$ from the jet ratio
independently of new physics as eq. (1).
If an exact determination of $\as(M_Z^2)$ from the jet ratio
still shows discrepancy with $\as$ from low energy value,
it may mean the exitence of other kinds of new physics different from
that described by eq. (1).

\acknowledgements

The work was supported
in part by the Korean Science and Engineering Foundation (KOSEF).

\newpage
\begin{table}
\caption{
Our global fit results to LEP observables in the Standard Model framework
and with the \non interactions.
}
\end{table}
%

%
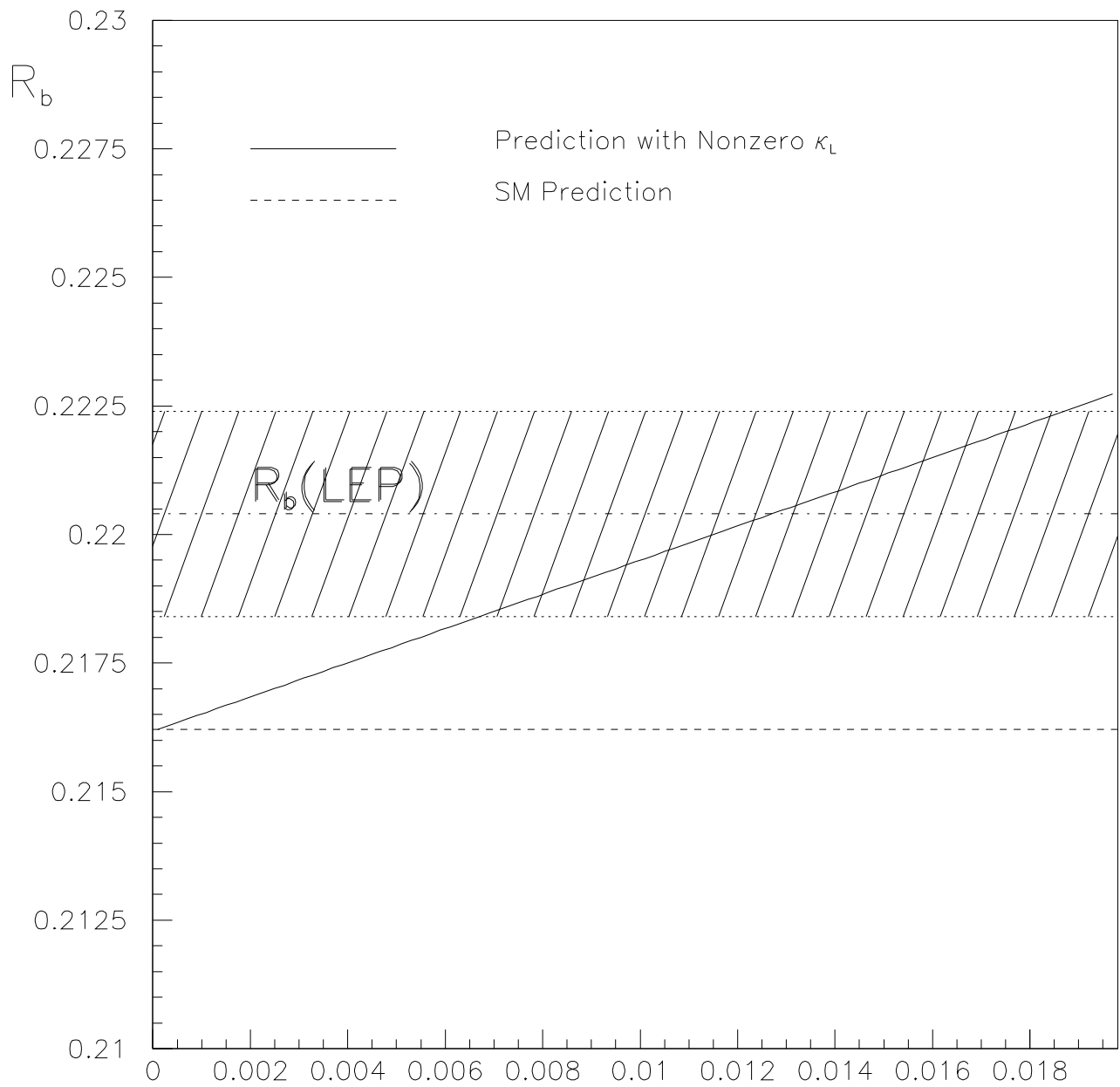
\begin{figure}
\caption{
Plot of $R_b$ as a function of the parameter $\kappa_L$.
The solid line denotes the prediction with nonzero $\kappa_L$
and the dashed line the Standard Model prediction.
The hatched area represents the LEP data within 1$\sigma$ error.
}
\end{figure}


\begin{references}
\bibitem{cdf} F. Abe et al., CDF Collaboration,
FERMILAB--PUB--94--116--E (1994).
\bibitem{lep} The LEP Electroweak Working Group, a Combination of
preliminary LEP Electroweak Results for the 1995 Winter Conferences,
preprint LEPEWWG/95--01, ALEPH 95--038, DELPHI 95--37, L3 Note 1736,
OPAL Note TN284.
\bibitem{ewgroup} The LEP Collaborations and the LEP Electroweak
Working Group, CERN/PPE/94--187.
\bibitem{non1} W. A. Bardeen, C. T. Hill and M. Lindner,
Phys. Rev. {\bf D 41}, 1647 (1990);
C. T. Hill and S. Parke, Phys. Rev. {\bf D 49}, 4454 (1994);
C. T. Hill, Phys. Lett. {\bf B 345}, 483 (1995).
\bibitem{non2} C. T. Hill and X. Zhang, Phys. Rev. {\bf D 51}, 3563 (1995).
\bibitem{shifman} M. Shifman, Mod. Phys. Lett. {\bf A 10}, 605 (1995).
\bibitem{kane} G. L. Kane, R. G. Stuart and J. D. Wells,
Preprint UM--TH--94--16, hep--ph/9505207.
\bibitem{zfit} D. Bardin et al., CERN--TH. 6443/92.
\bibitem{minuit} F. James and M. Roos,
Comp. Phys. Comm. {\bf 10}, 343 (1975).
\bibitem{jet} S. Bethke et al., Nucl. Phys. {\bf B 370}, 310 (1992);
Z. Kunszt, P. Nason, G. Marchesini and B. R. Webber,
Z physics at LEP 1, Vol 3, Event generators and software,
CERN yellow report 89--08 (1989).
\bibitem{bilenky} M. Bilenky, G. Rodrigo and A. Santamaria,
Nucl. Phys. {\bf B 439}, 505 (1995).
\end{references}
\end{document}